\documentclass[fp,twocolumn]{jpsj3} 
\usepackage{txfonts}
\usepackage{graphicx}  
\usepackage{bm}        
\usepackage[dvipdfmx]{color}
\usepackage{here}
\usepackage{color}
\usepackage{braket}
\usepackage{comment}
\graphicspath{{./figures_eps/}}

\newcommand{\tred}{}

\title{Ground-State Phase Diagram of the $S = 1$ One-Dimensional Kondo Lattice Model with a Uniaxial Anisotropy under Transverse Fields}

\author{Kohei Suzuki and 
Kazumasa Hattori
}
\inst{Department of Physics, Tokyo Metropolitan University, 
1-1 Minami-osawa, Hachioji, Tokyo 192-0397, Japan
}

\abst{ We study \tred{the} effects of transverse magnetic fields \tred{on} the $S=1$ one-dimensional Kondo lattice model with a uniaxial anisotropy \tred{using} the density matrix renormalization group.
The model can be regarded as a simplified one for analyzing \tred{the} Ising ferromagnetic superconductor URhGe.
We find various phases such as ferromagnetic and antiferromagnetic phases, Kondo plateau (KP) phases, \tred{Tomonaga--Luttinger liquids}, and \tred{fully polarized} phases.
In the KP phase, a \tred{pseudoplateau} emerges in the magnetization due to \tred{strongly bound} pairs between the local and the conduction electron spins. This is \tred{why we call} this phase \tred{the} Kondo plateau. 
At the critical field between the ferromagnetic and \tred{KP} phases, we find metamagnetic behavior in the magnetization curve. We discuss various correlation functions and the Friedel oscillations in \tred{detail}, and the experimental data under transverse magnetic fields in URhGe are discussed on the basis of the present results.
}

\begin{document}

\maketitle

\section{Introduction}
\tred{The} coexistence of ferromagnetism and superconductivity (SC) has attracted \tred{considerable} attention since the discovery in $\rm{UGe_{2}}$\cite{UGe2_FMS}. 
Since then, some other materials showing both ferromagnetism and SC have been \tred{found,} such as URhGe\cite{URhGe_FMS}, UIr\cite{UIr_FMS}, and UCoGe\cite{UCoGe_FMS}.
Among them, URhGe and UCoGe have been discussed in \tred{considerable detail} in recent years\cite{URhGe_RSC,URhGe_HTphadia,UCoGe_PTphadia,UCoGe_Hc2}. 
They show SC inside their ferromagnetic (FM) state in \tred{ambient} pressure\cite{URhGe_FMS,UCoGe_FMS}.
One of the common properties \tred{of} the U-based \tred{FM} superconductors is their strong Ising \tred{anisotropy,} and 
strongly anisotropic magnetic fluctuations have been observed in \tred{NMR} experiments\cite{URhGe_fluc1,URhGe_fluc2,UCoGe_fluc1,UCocGe_fluc2}.

For URhGe, the Ising \tred{FM} order takes place at \tred{$T_{\rm c}\sim 9.5\;{\rm K}$,} and the moment size is about $0.4\mu_{\rm B}$ along the c-axis. 
The SC emerges inside the FM state at $T_{\rm sc}=0.26\;{\rm K}$ and disappears for the magnetic field $H=2\;\rm{T}$ along the b-axis.
However, another SC emerges when the field increases further \tred{($H\sim 9\;\rm{T}$)}.\cite{URhGe_RSC}.
This is called \tred{reentrant} SC (RSC).
This RSC emerges near the critical magnetic field along the b-axis, above which the FM state disappears and the moment is polarized along the b-axis. 
Thus, the origin \tred{of} the RSC \tred{is considered to be} related to the physics of the transverse-field Ising model \tred{(fluctuation),} a well-known model that possesses a quantum criticality\cite{QPT_Book}. 
Detailed experimental investigations have revealed that the transition is \tred{first-order} and there is a tricritical point at \tred{a} finite temperature\cite{URhGe_TCP_TEP,URhGe_TCP_NMR}, 
while there \tred{is} a smooth change in the thermodynamic quantities such as the magnetization\cite{URhGe_metamag}. 
This suggests that it is a ``weak'' \tred{first-order} transition. 
Indeed, the magnetization strongly increases inside the RSC phase, which is typical metamagnetic behavior. 
This is also reflected in the relaxation time $T_{2}$ in the NMR experiment\cite{URhGe_fluc1}, in which $1/T_{2}$ shows strong enhancement.

\tred{Regarding} theoretical analyses, several works have been \tred{carried out}\cite{URhGe_Mineev1,URhGe_theor_XXZ,URhGe_Mineev2,URhGe_Mineev3,URhGe_Mineev4,UCoGe_theor,Lifshitz_trransition,Christopher}, which include \tred{studies on} SC mediated by spin-wave excitations near the critical field\cite{URhGe_theor_XXZ} and by phenomenological spin fluctuations\cite{UCoGe_theor}, and SC near Lifshitz transitions\cite{Lifshitz_trransition}. 
Inevitable line nodes in the superconducting gap functions characteristic \tred{of} the nonsymmorphic \tred{zigzag} chain structure along the a-axis \tred{have been} pointed out \tred{recently\cite{LineNode_UCoGe},} and also \tred{an} interesting topological SC for UCoGe \tred{has been} proposed\cite{Topological_SC_UCoGe}.

In this paper, we focus on the effects of transverse magnetic fields in the Kondo lattice model with strong Ising anisotropy.
To simplify the analysis, we consider the one-dimensional model and the local spin $S=1$.
The first simplification corresponds to ignoring inter-chain couplings in URhGe and related compounds.
The second one, taking \tred{the} $S=1$ model, enables us to discuss the uniaxial single-ion anisotropy \tred{while maintaing} the essential property of the U-based \tred{FM} SCs.
This is the lowest spin that has the uniaxial \tred{anisotropy,} and the analysis becomes much simpler than that for the larger spin.

For more than two decades, a number of studies about the $S=1/2$ Kondo lattice model
have been carried out\cite{Ueda_KLFerro,Assad,ODKL_Review,ODKL_GS,Otsuki,Smerat_KLQuasi}.
In the one-dimensional Kondo lattice model, it is well known that the FM state appears for \tred{the} strong Kondo coupling region\cite{Troyer_KLFerro,Tsunetsugu_KLFerro,Honner_Phase,Honner_Phase2,McCulloch_KLFerro}.
The existence of other FM phases \tred{has also been} discussed recently.\cite{Juozapavicius_KLFerro,McCulloch_KLFerro2,Garcia_KL_HL,Basylko_KLED,Peters_KLFerro}
In contrast, for larger spin models, including \tred{the} $S=1$ one, it has not been studied in \tred{detail} so far.
The $S=1$ Kondo lattice model has been analyzed \tred{on the basis of} mean-field theory\cite{ODKL_S1_1,ODKL_S1_2} \tred{to disucuss} U-based heavy fermions.
For impurity Kondo models, the Kondo model with an $S=1$ local spin interacting with $s=3/2$ conduction electrons has been discussed and shown to be an exotic non-Fermi liquid\cite{koga}.
Although much attention has been paid to the SC under magnetic fields in URhGe and UCoGe, there \tred{is} no study focusing on the ground-state properties and the phase diagram for microscopic models with strong Ising anisotropy in the Kondo lattice systems.
Thus, it is important to analyze the $S=1$ Kondo lattice model and to construct the phase diagram under transverse fields.
\tred{To} carry out the quantitative analysis, we employ the density matrix renormalization group (DMRG)\cite{DMRG}, which is a well-known powerful numerical tool for analyzing one-dimensional systems.

This paper is organized as follows. 
In Sect. 2, the model used in this study and some of \tred{the details of} the DMRG specific to the model are introduced. 
Then, we will show the numerical results, which include ground-state phase diagrams and various correlation functions in Sect. 3.
In Sect. 4, we will discuss our numerical results by comparing \tred{them} with experimental data in URhGe, and finally, \tred{we} will summarize the present results.

\section{Model}
In this section, we will introduce the model and explain \tred{its basic properties}. 
The Hamiltonian of \tred{the} $S=1$ one-dimensional Kondo lattice model with a uniaxial anisotropy $D$ under a transverse field $h$ is given as
\begin{align}
\hat{H}&=-t\sum_{j=1}^{N-1}\sum_{\sigma=\uparrow,\downarrow}\left(\hat{c}_{j\sigma}^{\dagger}\hat{c}_{j+1\sigma}+\hat{c}_{j+1\sigma}^{\dagger}\hat{c}_{j\sigma}\right) \nonumber \\
&+J\sum_{j=1}^{N}{\hat{\bm s}_{j}}\cdot{\hat{\bm{S}}}_{j}-h\sum_{j=1}^{N}\left(\hat{s}_{j}^{x}+\hat{S}_{j}^{x}\right)-D\sum_{j=1}^{N}\left(\hat{S}_{j}^{z}\right)^2.
\label{eq_ham}
\end{align}
Here, $N$ is the system size and $\hat{c}_{j\sigma}$ is the annihilation operator of the conduction electron at the $j$ site with the spin $\sigma=\uparrow \rm{or} \downarrow$. 
$\hat{\bm{s}}_{j}=(\hat{s}_{j}^{x},\hat{s}_{j}^{y},\hat{s}_{j}^{z})$ and $\hat{\bm{S}}_{j}=(\hat{S}_{j}^{x},\hat{S}_{j}^{y},\hat{S}_{j}^{z})$ are the spin operators for the conduction electron and the spin-1 local spin at the $j$ site,  respectively.
$J>0$ is the antiferromagnetic Kondo exchange \tred{coupling,} and the nearest-neighbor hopping $t$ is set to unity as a unit of energy.
\tred{To} represent the strong Ising anisotropy, we set $D=1$ for simplicity in this paper. 
Without loss of generality, we can assume that the transverse field $h$ is along \tred{the} $x$-direction.

In the presence of $h$, the \tred{$z$}-component of the total spins is not conserved.
It is important to choose the bases with which the Hamiltonian \tred{in Eq.} (\ref{eq_ham}) is block-diagonalized.
For this, we need to find the operators that commute with the Hamiltonian \tred{in Eq.} (\ref{eq_ham}).
The total number operator $\hat{N}_{c}=\sum_{i,\sigma}\hat{c}_{i\sigma}^{\dagger}\hat{c}_{i\sigma}$ and the total spin inversion operator $\hat{P}=\prod_{j=1}^{N}\hat{P}_{j}$ meet this requirement.
The local spin inversion operator $\hat{P}_{j}$ acts on the local bases for the conduction electrons as follows:
\begin{align}
&\hat{P}_{j}\ket{\rm vac}_{j}=\ket{\rm vac}_{j}, \;\hat{P}_{j}\ket{\uparrow}_{j}=\ket{\downarrow}_{j}, \nonumber \\
&\hat{P}_{j}\ket{\downarrow}_{j}=\ket{\uparrow}_{j}, \;\hat{P}_{j}\ket{\uparrow\downarrow}_{j}=\ket{\downarrow\uparrow}_{j},
\end{align}
where $\ket{\rm{vac}}_{j}$ indicates the empty state for the $j$ site, $\ket{\sigma}_{j}=\hat{c}_{j\sigma}^{\dagger}\ket{\rm{vac}}_{j}$, and $\ket{\uparrow\downarrow}_{j}=\hat{c}_{j\uparrow}^{\dagger}\hat{c}_{j\downarrow}^{\dagger}\ket{\rm{vac}}_{j}$.
For the local spins,
\begin{align}
\hat{P}_{j}\ket{\Uparrow}_{j}=\ket{\Downarrow}_{j},\;\hat{P}_{j}\ket{0}_{j}=\ket{0}_{j},\;\hat{P}_{j}\ket{\Downarrow}_{j}=\ket{\Uparrow}_{j},
\end{align}
where the local bases for the local spins are the eigenstates for $\hat{S}_{j}^{z}$: $\ket{\Uparrow}_{j}$, $\ket{0}_{j}$, and $\ket{\Downarrow}_{j}$ with the \tred{eigenvalues} $1$, $0$, and $-1$, respectively. 
The eigenvalues of $\hat{P}_{j}$ are \tred{$P_{j}=\pm 1$ since $\hat{P}_{j}^{2}=1$}.
The eigenstates of $\hat{P}_{j}$ for the conduction electrons are summarized as
 \begin{align}
 &\ket{\rm vac}_{j}\;\cdots\; P_{j}=1,\nonumber \\
 &\hat{c}_{j,\rm e}^{\dagger}\ket{\rm vac}_{j}:=\frac{1}{\sqrt{2}}(\ket{\uparrow}_{j}+\ket{\downarrow}_{j})\;\cdots\; P_{j}=1,\nonumber \\
 &\hat{c}_{j,\rm o}^{\dagger}\ket{\rm vac}_{j}:=\frac{1}{\sqrt{2}}(\ket{\uparrow}_{j}-\ket{\downarrow}_{j})\;\cdots\; P_{j}=-1,\nonumber \\
 &\hat{c}_{j,\rm o}^{\dagger}\hat{c}_{j,\rm e}^{\dagger}\ket{\rm vac}_{j}:=\ket{\uparrow\downarrow}_{j}\;\cdots\; P_{j}=-1.
 \end{align}
Here, $\hat{c}_{j,\rm e(o)}^{\dagger}$ is the creation operator for the conduction electron with even (odd) parity.
They are defined as
 \begin{align}
 \hat{c}_{j,\rm e}^{\dagger}=\frac{1}{\sqrt{2}}(\hat{c}_{j\uparrow}^{\dagger}+\hat{c}_{j\downarrow}^{\dagger}),\;
  \hat{c}_{j,\rm o}^{\dagger}=\frac{1}{\sqrt{2}}(\hat{c}_{j\uparrow}^{\dagger}-\hat{c}_{j\downarrow}^{\dagger}).
 \end{align}
\tred{The} eigenstates of $\hat{P}_{j}$ for the local spins \tred{are given as}
\begin{align}
&\ket{\rm EV}_{j}:=\frac{1}{\sqrt{2}}(\ket{\Uparrow}_{j}+\ket{\Downarrow}_{j})\;\cdots\; P_{j}=1,\nonumber \\
&\ket{0}_{j}\;\cdots\; P_{j}=1,\nonumber \\
&\ket{\rm OD}_{j}:=\frac{1}{\sqrt{2}}(\ket{\Uparrow}_{j}-\ket{\Downarrow}_{j})\;\cdots\; P_{j}=-1.
\end{align}
With these bases, the Hamiltonian \tred{in Eq.} (\ref{eq_ham}) is rewritten as
\begin{align}
\hat{H}&=-t\sum_{j=1}^{N-1}\sum_{p=\rm e,o}\left(\hat{c}_{j,p}^{\dagger}\hat{c}_{j+1,p}+\hat{c}_{j+1,p}^{\dagger}\hat{c}_{j,p}\right)\nonumber \\
&+J\sum_{j=1}^{N}\left(\hat{s}_{j}^{x}\hat{S}_{j}^{x}+\hat{s}_{j}^{y}\hat{S}_{j}^{y}+\hat{s}_{j}^{z}\hat{S}_{j}^{z}\right)-h\sum_{j=1}^{N}\left(\hat{s}_{j}^{x}+\hat{S}_{j}^{x}\right)\nonumber \\
&-D\sum_{j=1}^{N}\left(1-\ket{0}_{j}\bra{0}_{j}\right),
\label{eq_ham2}
\end{align}
and the spin operators are represented in the new bases as
\begin{align}
&\hat{s}_{j}^{x}=\frac{1}{2}\left(\hat{c}_{j,\rm e}^{\dagger}\hat{c}_{j,\rm e}-\hat{c}_{j,\rm o}^{\dagger}\hat{c}_{j,\rm o}\right),\\
&\hat{s}_{j}^{y}=\frac{i}{2}\left(\hat{c}_{j,\rm e}^{\dagger}\hat{c}_{j,\rm o}-\hat{c}_{j,\rm o}^{\dagger}\hat{c}_{j,\rm e}\right),\\
&\hat{s}_{j}^{z}=\frac{1}{2}\left(\hat{c}_{j,\rm e}^{\dagger}\hat{c}_{j,\rm o}+\hat{c}_{j,\rm o}^{\dagger}\hat{c}_{j,\rm e}\right),\\
&\hat{S}_{j}^{x}=\ket{\rm EV}_{j}\bra{0}_{j}+\ket{0}_{j}\bra{\rm EV}_{j},\\
&\hat{S}_{j}^{y}=i\left(\ket{0}_{j}\bra{\rm OD}_{j}-\ket{\rm OD}_{j}\bra{0}_{j}\right),\\
&\hat{S}_{j}^{z}=\ket{\rm EV}_{j}\bra{\rm OD}_{j}+\ket{\rm OD}_{j}\bra{\rm EV}_{j}.
\end{align}
The DMRG calculations have been \tred{carried out} in these bases in each of the subspaces specified by the electron number $N_{c}$ and the total spin parity $P=\pm 1$ (the eigenvalue of $\hat{P}$).

Before closing this section, let us comment on the particle-hole symmetry.
\tred{The} Hamiltonian \tred{in Eq.} (\ref{eq_ham2}) is invariant under the particle-hole \tred{transformation}
\begin{align}
\hat{c}_{j,\rm e}\rightarrow(-1)^{j}\hat{c}_{j,\rm o}^{\dagger},\; \hat{c}_{j,\rm o}\rightarrow(-1)^{j+1}\hat{c}_{j,\rm e}^{\dagger}.
\label{eq_pht}
\end{align}
The filling of the conduction electrons $n_{c}$ changes under the transformation as follows:
\begin{align}
n_{c}=\frac{N_{c}}{N}=\frac{1}{N}\sum_{j=1}^{N}\sum_{p=\rm e,o}\braket{\hat{c}_{j,p}^{\dagger}\hat{c}_{j,p}}\rightarrow2-n_{c},
\label{eq_fill}
\end{align}
where $\braket{\hat{A}}$ represents the ground-state expectation value of the operator $\hat{A}$.
Considering Eq. (\ref{eq_fill}), one can easily find that systems with the filling $n_{c}$ are equivalent to those for $2-n_{c}$ and it is sufficient to consider $n_{c}\leq 1$.

Since the parity index for the conduction electrons changes through Eq. (\ref{eq_pht}), 
the total spin parity for the electrons depends on the electron number $N_{c}$.
It is invariant for $N_{c}$ being even, while it changes for odd \tred{$N_{c}$}.
Since the particle-hole transformation does not affect the parity for the local spins, 
the total spin parity is invariant if $N_{c}$ is even, while it changes if $N_{c}$ is odd.
Thus, systems for $n_c>1$ can be trivially mapped to those for $n_c<1$ and it is sufficient to consider $n_{c}\leq 1$.

\section{\tred{Results}}
In this section, we will show the numerical results in \tred{detail}.
DMRG calculations have been \tred{carried out} for $N=32,64,96,128$ and \tred{with} at least $m=200$ states \tred{kept} in the truncation procedures.
We will mainly discuss the results for $N=128$.
If necessary, the $N$ and $m$ \tred{dependences are} discussed.
In the following, we will first show the $h$-$J$ phase diagram for $n_{c}=0.5$ and then discuss the physical quantities such as the magnetizations, the correlation functions, and the critical properties.
For other values of $n_c$, the results are discussed in Sect. 3.5, in which qualitatively similar phase diagrams are obtained with some new phases.

\subsection{Phase diagram}

\begin{figure}[t]
	\begin{center}
	 \includegraphics[scale=0.3]{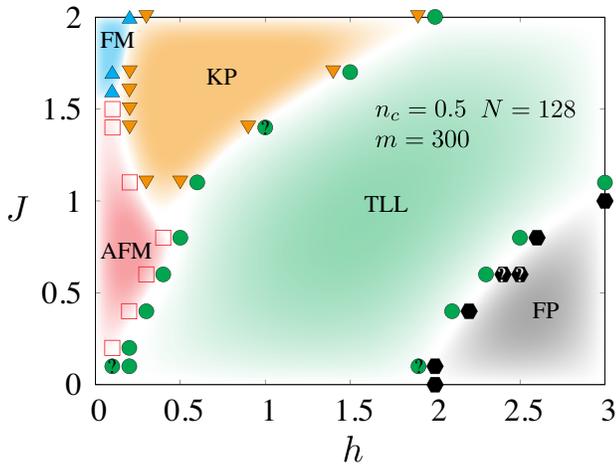}
	 \caption{(Color online) Ground-state $h$-$J$ phase diagram for $N=128$ and $n_{c}=0.5$.
Each phase is identified by the long-distance behavior of the spin-spin correlation functions $\chi^{sc}_{z}(r)$ and $\chi_{z}^{S}(r)$. They do not decay and are finite in the Ising-ordered phases, while \tred{they} decay exponentially in the KP and FP phases. In the TLL phase, they show a \tred{power-law} decay.
Symbols with ``?'' near phase boundaries indicate that the ground state is not identified within the system sizes $N \leqq 128$ and the cutoff $m=300$.
}
		\label{phase-diagram}
	\end{center}
\end{figure}

\begin{figure}[t]
	\begin{center}
		\includegraphics[scale=0.25]{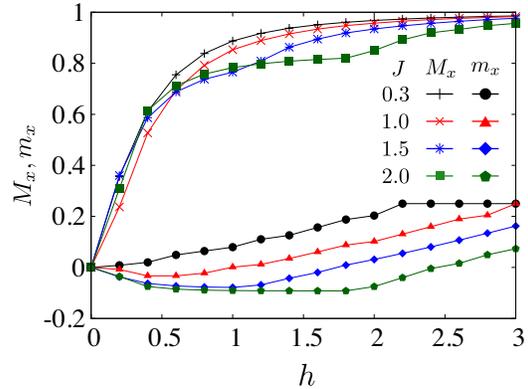}
		\caption{(Color online) $h$ \tred{dependences} of $M_{x}$ and $m_{x}$ for $J=0.3$, $1.0$, $1.5$, and $2.0$, and $N=128$, $n_{c}=0.5$ with the cutoff $m=200$.
		$M_{x}$ \tred{represents} the local spins and $m_{x}$ \tred{represents} the conduction electrons.
		See Eq. (\ref{eq_magnetization}).
		}
		\label{mean_Sx1_Sx0_e64}
	\end{center}
\end{figure}

Let us first show the ground-state $h$-$J$ phase diagram determined by the DMRG method for the typical conduction electron filling $n_c=0.5$.
Here, we will explain \tred{the} overall features and properties of phases appearing in the diagram and will discuss various physical quantities later in this section.

Figure \ref{phase-diagram} shows the $h$-$J$ phase diagram for $n_c = 0.5$. 
There are various phases such as \tred{FM}, antiferromagnetic (AFM), the \tred{Tomonaga--Luttinger} liquid (TLL), Kondo plateau (KP) (we will explain the meaning later), and \tred{fully polarized} (FP) phases.
Each of the phases shows a characteristic long-distance asymptotic dependence in the spin-spin correlation functions such as 
\begin{align}
\chi^S_z(r):=\braket{\hat{S}^{z}_{r+N/2}\hat{S}^{z}_{N/2}}, \quad \chi^{sc}_z(r):=\braket{\hat{s}^{z}_{r+N/2}\hat{s}^{z}_{N/2}}.
\label{chiSc}
\end{align} 
The FM and \tred{AFM} orders are \tred{Ising-like}, which are possible even in one-dimensional 
systems at zero temperature.
These phases appear for small $h$. In particular, the FM phase emerges for large $J$ and small $n_c$ (see also Sect. 3.5).
The spin-spin correlation functions do not vanish as $r\rightarrow\infty$ in these phases, which is the signature of  \tred{symmetry breaking}.

In the KP phase, \tred{tightly bound} local Kondo ``doublets'' are formed by the conduction electrons and the local \tred{spins,} 
and one of the two components gains the magnetic energy due to $h$.
Note that the magnitudes of the local spin and the conduction electron \tred{spins} are different in our model. 
\tred{Destroying} such a \tred{tightly bound} doublet \tred{requires} a large magnetic field $\sim J$.
This is reflected in the \tred{pseudoplateau} in the magnetization (Fig. \ref{mean_Sx1_Sx0_e64}) and in the local spin-spin correlation (Fig. \ref{mean_S0S1_e64}):
\begin{align}
\braket{\hat{\bm s}\cdot \hat{\bm S}}:=\frac{1}{N}\sum_{j=1}^N \braket{\hat{\bm s}_j\cdot \hat{\bm S}_j}.
\label{s_dot_S}
\end{align}
We call this phase \tred{the} ``Kondo plateau'' after these observations. 
As will be discussed in Sect. 3.3, the spin-spin correlation functions \tred{[Eq. (\ref{chiSc})]} decay exponentially for $r\rightarrow\infty$, which indicates \tred{that} the spin excitations have a finite gap.

In a wide range of \tred{the} parameter space, \tred{the TLL phase appears}, where the spin-spin correlation functions exhibit power-law decay, which indicates \tred{quasi-long-range} order. 
Strictly speaking, it is ``TLL'' for the charge sector in other phases, but \tred{here, we} use the name TLL in order to distinguish this phase and others by focusing on the spin sector only. 
For sufficiently high fields, the FP state appears and this phase is a trivial one expected in the band-shift picture due to the external magnetic field $h$.

\subsection{Magnetization curve and Kondo correlations}\label{sec:3.2}
Now, we discuss the response of the ground state against the external transverse magnetic field $h$.
One of the most relevant quantities for experimental studies is the \tred{magnetization,} and let us \tred{discuss this}.

Figure \ref{mean_Sx1_Sx0_e64} shows the magnetizations along the $x$-direction,
\begin{align}
M_x=\frac{1}{N}\sum^N_{j=1}\braket{\hat{S}^{x}_{j}}, \quad m_x=\frac{1}{N}\sum^N_{j=1}\braket{\hat{s}^{x}_{j}},
\label{eq_magnetization}
\end{align}
as a function of $h$ for several values of $J$ for $n_c=0.5$.
$M_x$ is induced as $h$ \tred{increases,} and around $h\sim 3.0$, the moment reaches the saturated value 1 for all \tred{$J$}. 
\tred{Note} that for larger $J$, $J = 2.0$ and $J=1.5$, $M_x$ exhibits \tred{nonmonotonic} $h$ dependence.
For example, for $J=2.0$, below $h=h_{\rm KP}\simeq 1.92$, $M_x$ monotonically increases as if the full moment were $\sim 0.8$ and exhibits (almost) no $h$ dependence 
between $h\sim 1$ and $h_{\rm KP}$.
Correspondingly, $m_x$ decreases for small $h$ and is almost constant with a negative value.
In contrast, for $J=0.3$, it increases monotonically and reaches the saturated value $1/4$ at $h\simeq 2.2$.
Note that $m_x$ for large $J$ \tred{also} starts to increase at $h=h_{\rm KP}$.

\begin{figure}[t]
	\begin{center}
		\includegraphics[scale=0.25]{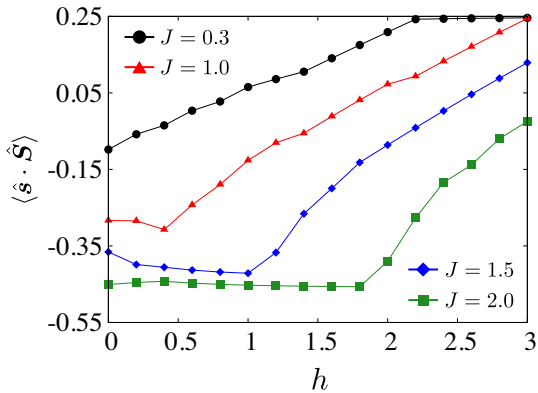}
		\caption{(Color online) $h$ dependence of $\braket{\hat{\bm s}\cdot \hat{\bm S}}$, the correlation between the local spins and the electron spins, for $J=0.3$, $1.0$, $1.5$, and $2.0$, and $N=128$, $n_{c}=0.5$ with the cutoff $m=200$.
		}
		\label{mean_S0S1_e64}
	\end{center}
\end{figure}

\begin{figure}[t]
	\begin{center}
		\includegraphics[scale=0.25]{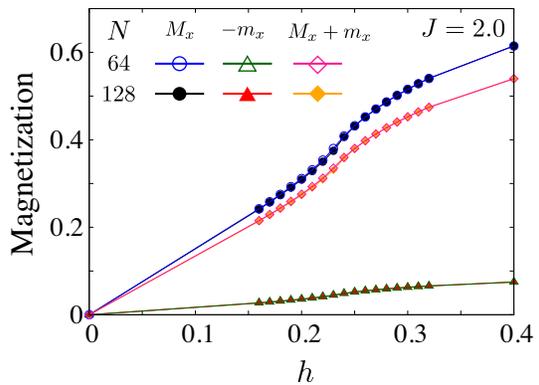}
		\caption{(Color online) 
		Magnetizations $M_{x}$ and $m_{x}$ as \tred{functions} of $h$ near the phase boundary between the FM and \tred{KP} phases for $J=2.0$, $N=64$, $128$, and $n_{c}=0.5$ with the cutoff $m=200$. Note that $-m_{x}$ (triangles) is plotted for the conduction electron.
		 }
		\label{J20_mean_Sx1_Sx0_e64_FMTOKP}
	\end{center}
\end{figure}

\tred{To} understand this \tred{nonmonotonic} $h$ dependence, we analyze the correlation between the conduction electron spins and the local spins: 
$\braket{\hat{\bm{s}}\cdot\hat{\bm{S}}}$ [Eq. (\ref{s_dot_S})]. Figure \ref{mean_S0S1_e64} shows the $h$ dependence of $\braket{\hat{\bm{s}}\cdot\hat{\bm{S}}}$. 
For $J=2.0$, $\braket{\hat{\bm{s}}\cdot\hat{\bm{S}}}\simeq -0.45$, and this means \tred{that} the electron and the local spin 
are almost frozen, forming an \tred{antiparallel} configuration, which can \tred{also be seen} in Fig. \ref{mean_Sx1_Sx0_e64}. Since the magnitudes of the two are different, a residual moment \tred{remains}.
This residual moment aligns along the external field $h$, leading to the positive $M_x$ and the negative 
$m_x$ in Fig. \ref{mean_Sx1_Sx0_e64}, since the residual moment is parallel to the local spin.
This indicates that the plateau in $M_x$ is due to the strong Kondo correlations; the ``spin molecule'' is formed between the conduction electron and the local spin. 
Above $h_{\rm KP}$, this molecule is resolved and $M_x$, $m_x$, and $\braket{\hat{\bm{s}}\cdot\hat{\bm{S}}}$ start to increase.
The transition at $h=h_{\rm KP}$ separates the KP and \tred{TLL} phases. 
Unfortunately, it is \tred{difficult} to judge whether the transition is first- or second-order in the present calculations for $N=128$ and $m=500$, 
although there are some indications \tred{that it is} the \tred{second-order}; 
in the KP phase, the correlation lengths of the spin-spin correlation functions seem to diverge toward the phase boundary at $h=h_{\rm{KP}}$, 
but their $m$ dependence does not saturate.
\tred{$M_{x}$, $m_{x}$, and $\braket{\hat{\bm{s}}\cdot\hat{\bm{S}}}$ exhibit} abrupt changes without discontinuity at $h=h_{\rm{KP}}$ 
when their $N$ dependence is analyzed. 
For the TLL side, the spin-spin correlation functions show power-law decay, which means \tred{that} the correlation length is infinite.

At low fields, there is another phase transition that separates the KP and \tred{Ising-ordered} phases for $J=2.0$ and 1.5. 
Figure \ref{J20_mean_Sx1_Sx0_e64_FMTOKP} shows the low-field magnetization curve near the transition. The phase boundary corresponds to the field where \tred{$M_x$} shows a (\tred{metamagnetic-like}) steep increase as a function of the field; $h_c \sim 0.24$ for $J=2.0$. Although 
it is not \tred{direct} evidence for the phase transition, it suggests \tred{that} there is some anomaly at $h=h_c$. 
We will discuss this point in Sect. \ref{sec:3.3} by analyzing correlation \tred{functions,} and the low-field phase 
turns out to be an Ising FM state for $J=2.0$ and 1.5. For smaller $J=0.3$, there is no KP phase and there is one 
transition from the Ising AFM to \tred{TLL} phases as $h$ increases. For $J=1.0$, it seems \tred{that} there is a 
small region of the KP phase between the FM and \tred{TLL} phases around $h=0.4$ (see Figs. \ref{mean_Sx1_Sx0_e64} and \ref{mean_S0S1_e64}).

\begin{figure}[tb]
	\begin{center}
		\includegraphics[scale=0.25]{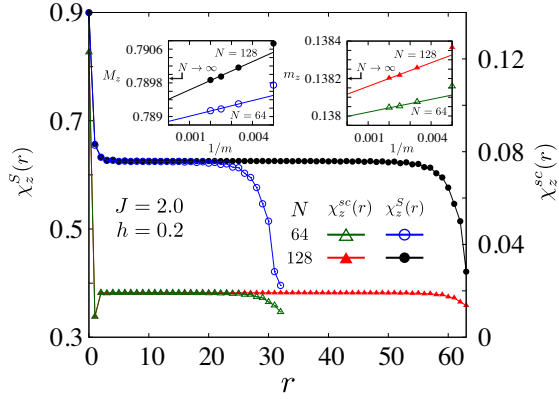}
		\caption{(Color online) 
		$r$ dependence of the spin-spin correlation functions $\chi_{z}^{S}(r)$ and $\chi_{z}^{sc}(r)$ for $N=64$ and $128$ inside the FM phase: $J=2.0$, $h=0.2$, and $n_{c}=0.5$ with the cutoff $m=200$.
		The suppression for $r\sim N/2$ is due to the boundary effects. 
		Inset: \tred{$m$ dependences} of the magnetizations $M_{z}$ and $m_{z}$. 
		Here, $M_{z}^{2} = \chi_{z}^{S}(N/4)$ and $m_{z}^{2} = \chi_{z}^{s}(N/4)$.
		The lines represent the linear fits by using the \tred{data} for $m=300, 400$, and 500. 
		The arrows indicate the magnetizations extrapolated to $1/N\rightarrow 0$ by using the extrapolated values $1/m\rightarrow 0$ for $N=64$ and 128, which remain finite.
		}
		\label{J20h02_Sz1_CF_Sz0_CF}
	\end{center}
\end{figure}

\subsection{Correlation functions}\label{sec:3.3}
As discussed above, there are various phases as functions of $J$ and $h$. 
\tred{To} clarify the nature of each phase, analyzing the correlation functions, e.g., Eq. (\ref{chiSc}), is important.
One can unveil the spatial modulation and the long-distance asymptotic behavior by examining appropriate correlation functions. 
In our calculations, the cutoff $m$ is kept up to 500 and the system size $N$ to 128. 
In the following, we will show the $m$ and $N$ \tred{dependences} of the correlation functions in the \tred{Ising-ordered} phases and discuss the finite size effects. 
\tred{In} the other phases, \tred{they are very small}.

The Ising-ordered phases are stabilized for small $h$.
When $J$ is large, the FM phase appears and $\chi^S_z(r)$ and $\chi^{sc}_z(r)$ [Eq. (\ref{chiSc})] exhibit long-range correlations.  
In the FM phase, the magnetization along the $z$-direction is estimated to be $M_z^2=\lim_{r\to \infty} \chi^S_z(r)$ and $m_z^2=\lim_{r\to \infty}\chi^{sc}_z(r)$.
Figure \ref{J20h02_Sz1_CF_Sz0_CF} shows the $r$ dependence of $\chi^S_z(r)$ and $\chi^{sc}_z(r)$
for $J=2.0$ and $h=0.2$, i.e., in the FM phase. 
One can see that  $\chi^S_z(r)$ and $\chi^{sc}_z(r)$ do not decay for \tred{a} long distance except for the effect of the open boundary, 
which indicates \tred{that} the system is in the FM state. 
The inset of Fig. \ref{J20h02_Sz1_CF_Sz0_CF} shows the cutoff $m$ dependence of the magnetizations $M_z^2=\chi^S_z(N/4)$ and $m_z^2=\chi^{sc}_z(N/4)$ for $N=64$ and 128.
Here, we replace $r\rightarrow\infty$ by $r=N/4$.
Extrapolating them to $1/m\rightarrow 0$ \tred{and} then $1/N\rightarrow 0$, we obtain $M_z\sim 0.79$ and $m_z\sim 0.14$. 
Thus, this analysis confirms that the system is in the FM phase.

Let us now discuss the nature of this FM phase.
As shown in Fig. \ref{mean_S0S1_e64}, 
$\braket{\hat{\bm{s}}\cdot\hat{\bm{S}}}$ is negative, which means that the local spins and 
conduction electron \tred{spins} are \tred{antiparallel} with each other
\tred{owing} to the strong Kondo exchange coupling $J$. This kind of FM phase was also 
found in \tred{the} $S = 1/2$ Kondo lattice model.\cite{Troyer_KLFerro,Tsunetsugu_KLFerro,Honner_Phase,Honner_Phase2} 
\tred{In order for the conduction electrons to gain the kinetic energy}, 
the conduction electrons hop to the neighboring 
sites with their spins antiparallel to the local ones\cite{double-exchange}. 
Actually, the FM phase is stabilized for small filling, 
since the conduction electrons \tred{can easily} traverse. See also Sect. \ref{sec:3.5} for the phase diagrams for $n_{c}\neq0.5$.

\begin{figure}[t]
	\begin{center}
		\includegraphics[scale=0.28]{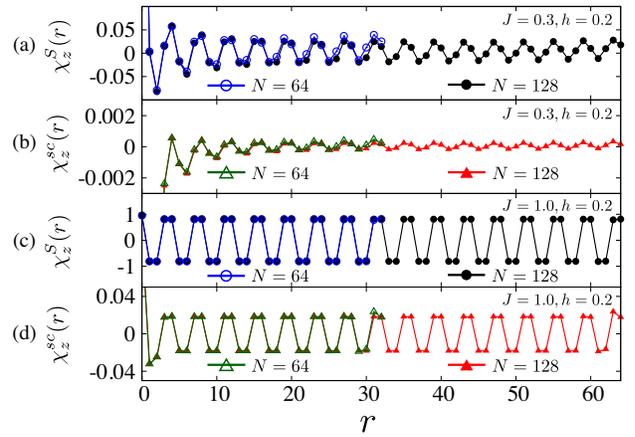}
		\caption{(Color online) 
		$r$ dependence of the spin-spin correlation functions for $N=64$ (open symbols), $128$ (filled symbols), and $n_{c}=0.5$ with the cutoff $m=200$: (a) $\chi_{z}^{S}(r)$ for $(J, h)=(0.3, 0.2)$, (b) $\chi_{z}^{sc}(r)$ for $(J, h)=(0.3, 0.2)$,  
		(c) $\chi_{z}^{S}(r)$ for $(J, h)=(1.0, 0.2)$, and (d) $\chi_{z}^{sc}(r)$ for $(J, h)=(1.0, 0.2)$.
		One can see \tred{an} ``up-up-down-down'' structure.
		}
		\label{J10h02J03h02_Sz1_CF_Sz0_CF}
	\end{center}
\end{figure}

\begin{figure}[t]
	\begin{center}
		\includegraphics[scale=0.37]{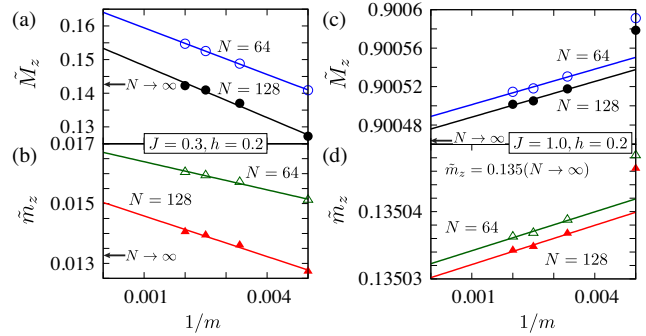}
		\caption{(Color online) 
		\tred{Cutoff} $m$ dependence of the AFM moments $\tilde{M}_z$ and $\tilde{m}_z$ for $N=64$ (open symbols) and $128$ (filled symbols) inside the AFM phases with $n_{c}=0.5$: 
		(a) $\tilde{M}_z$ for $(J,h)=(0.3, 0.2)$, (b) $\tilde{m}_z$ for $(J,h)=(0.3, 0.2)$, (c) $\tilde{M}_z$ for $(J,h)=(1.0, 0.2)$, and (d) $\tilde{m}_z$ for $(J,h)=(1.0, 0.2)$. 
		 The lines represent the linear fits by using the \tred{data} for $m=200, 300, 400$, and 500 for $(J,h)=(0.3, 0.2)$ and $m=300, 400$, and $500$ for $(J,h)=(1.0, 0.2)$.
		}
		\label{J10h02J03h02_Sz1_CF_Sz0_CF_mdep}
	\end{center}
\end{figure}

For small $J$ and $h$, the AFM phase appears. Figures \ref{J10h02J03h02_Sz1_CF_Sz0_CF}(a) and \ref{J10h02J03h02_Sz1_CF_Sz0_CF}(b) show the $r$ dependence of 
$\chi^S_z(r)$ and $\chi^{sc}_z(r)$, respectively, for $J = 0.3$ and $h = 0.2$. $\chi_z^S(r)$ shows \tred{a four-site-periodicity} structure and no decay for large $r$. 
This structure becomes clearer and shows \tred{an} ``up-up-down-down'' structure as $J$ increases as shown in Figs. \ref{J10h02J03h02_Sz1_CF_Sz0_CF}(c) and \ref{J10h02J03h02_Sz1_CF_Sz0_CF}(d) for $J=1.0$. 
Figure \ref{J10h02J03h02_Sz1_CF_Sz0_CF_mdep} shows the cutoff $m$ dependence of the AFM moments for $N=64$ and $128$.
Here, the AFM moments $\tilde{M}_z^2$ and $\tilde{m}_z^2$ are defined as the average values of $|\chi^{S,sc}_z(r)|$ from $r=N/4-1$ to $N/4+2$.
The AFM moments in the thermodynamic limit are estimated to be $(\tilde{M}_z, \tilde{m}_z)\sim (0.14, 0.013)$ for $(J,h)=(0.3, 0.2)$ and $(\tilde{M}_z, \tilde{m}_z)\sim (0.9, 0.13)$ for $(J,h)=(1.0, 0.2)$.

The main origin \tred{of} this AFM state is expected 
to be the \tred{Ruderman–Kittel–Kasuya–Yosida (RKKY)} interactions.\cite{RKKY1,RKKY2,RKKY3} When the Kondo interaction $J$ is small, the $J$ term can 
be regarded as a perturbation, and \tred{thus} the lowest-order interaction between the 
local spins is the RKKY interaction mediated by conduction electrons. 
To simplify the discussion, consider the weak-coupling limit. 
Since $D = 1\gg J$, the large $D$ prohibits the local spin from flipping by $\pm 1$ and it 
leads to the Ising-like interaction between the local spins $J_{\rm eff}$ \tred{given by} 
\begin{align}
&J_{\rm eff}(q)\sim  
\left(J+\frac{J^{2}}{2D}\right)^{2} \chi^{sc}_{z0}(q).
\label{effective_int}
\end{align}
Here, $\chi^{sc}_{z0}(q)$ represents the $z$ component of the static spin susceptibility with the wavenumber $q$
 for the \tred{noninteracting} conduction electrons.
The dominant Fourier component of \tred{Eq. (\ref{effective_int})} is \tred{the} one with $q=2k_{\rm F}^{0}$, where $k_{\rm F}^{0}$ is the Fermi wavenumber for $J=h=0$.
Thus, it is natural to expect that the period of the oscillation in $\chi^S_z(r)$ corresponds to the inverse of $2k_{\rm F}^{0}$. Indeed, when $n_c = 1/2$ and $h=0$, the wavelength 
is estimated to be $2\pi /(2k_{\rm F}^{0}) \sim 2/n_c = 4$, which is consistent with the ``up-up-down-down'' 
structure \tred{shown} in Fig. \ref{J10h02J03h02_Sz1_CF_Sz0_CF}. 
Since $\chi_{z}^{S,sc}(r)$ is related to the \tred{interband} spin fluctuations under a finite $h$, this argument (by setting $h=0$) is qualitatively valid for $h>0$.
As will be discussed in Sect. 3.5, $\chi^{S,sc}_z(r)$ has another structure corresponding to $6k_{\rm F}^{0}$ for other fillings.

Now, we discuss the phases with a gap in the spin sector.
In general, when a target sector possesses a finite excitation gap, the corresponding correlation 
function $\chi(r)$ decays exponentially as $r\to \infty$. The magnitude of the gap is characterized 
by the correlation length $\xi$ defined by $\chi(r) \sim \exp(-r/\xi)$. 
In our phase diagram \tred{(Fig. \ref{phase-diagram})}, there are three types of such gapped states. 
One is the FP phase, which is a trivial state for high magnetic fields. 
The second is the KP phase and the third is the Ising-ordered FM and AFM phases as discussed previously.

\begin{figure}[t]
	\begin{center}
		\includegraphics[scale=0.15]{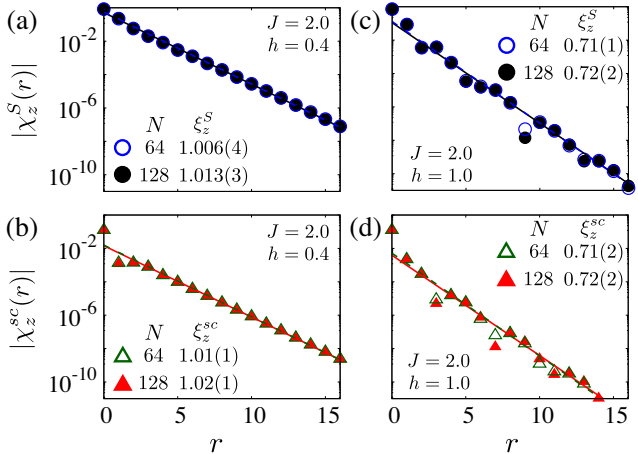}
		\caption{(Color online) 
		$r$ dependence of the spin-spin correlation functions for $N=64$ (open symbols), 128 (filled symbols), and $n_{c}=0.5$ with the cutoff $m=200$ in the KP phase: (a) $|\chi^{S}_{z}(r)|$ for $(J, h)=(2.0, 0.4)$, (b) $|\chi^{sc}_{z}(r)|$ for $(J, h)=(2.0, 0.4)$, 
		(c) $|\chi^{S}_{z}(r)|$ for $(J, h)=(2.0, 1.0)$, and (d) $|\chi^{sc}_{z}(r)|$ for $(J, h)=(2.0, 1.0)$.
		The (dashed) lines represent the fit by $A\exp(-r/\xi_{z}^{S,sc})$ for $N=128\;(64)$, 
		where $A$ and $\xi_{z}^{S,sc}$ are the fitting parameters.
		}
		\label{J20h10_Sz1_CF_Sz0_CF}
	\end{center}
\end{figure}

Let us first discuss \tred{the FP phase}. 
When $h$ is much larger than all the other parameters $t, J$, and $D$, 
the lower conduction electron band split by $h$ is filled with $N_c$ electrons 
(we are considering the case with $n_c\le 1$). 
Flipping a single spin generates \tred{interband} excitations and \tred{requires} a finite energy $\sim h$. 
Thus, since $\chi^{sc}_z(r)$ measures the transverse spin excitations with respect to the direction 
parallel to $h$ in the $x$-direction, $\chi^{sc}_z(r)$ should decay exponentially.

\tred{For the KP phase}, 
the \tred{directions} of the local and \tred{electron} spins \tred{are} opposite 
and they form spin molecules due to the large $J$ as discussed in Sect. \ref{sec:3.2}. Thus, flipping 
either local or conduction electron spins \tred{requires} a finite energy $\Delta E\sim J$ corresponding to the bound-state energy for the tightly bound spins. 
This means that the ground-state wavefunction is well approximated by the direct product state of local ones. 
For the single-site (i.e., $N=1$) model with $N_e=1$, there is a level crossing for $h\sim J$; 
the Kondo doublet-like state is destabilized for large magnetic fields. 
This is indeed consistent with our DMRG results \tred{and shows} a phase transition between the KP and \tred{TLL} phases for $h\sim J$.
Figures \ref{J20h10_Sz1_CF_Sz0_CF}(a) and \ref{J20h10_Sz1_CF_Sz0_CF}(b) [\ref{J20h10_Sz1_CF_Sz0_CF}(c) and \ref{J20h10_Sz1_CF_Sz0_CF}(d)] show the $r$ dependence of $\chi^S_z(r)$ and $\chi^{sc}_z(r)$, respectively, for $(J, h)=(2.0, 0.4)$ $[(2.0, 1.0)]$ in the KP phase.
The linear decay in the \tred{semilog} plot means that $\chi^S_z(r)$  and $\chi^{sc}_z(r)$ decay exponentially. 
By fitting the data, we obtain the correlation lengths $\xi^S_z$ and $\xi^{sc}_z$; 
they are similar and $\xi^S_z\simeq \xi^{sc}_z\sim 1$ for these parameter sets. This is naturally understood by noting that 
the correlation lengths are determined by \tred{the ``strength'' of the spin molecules}. 
These short correlation lengths indicate that the KP phase is well described by the direct product of the local wavefunction as discussed above.

\begin{figure}[t]
	\begin{center}
		\includegraphics[scale=0.15]{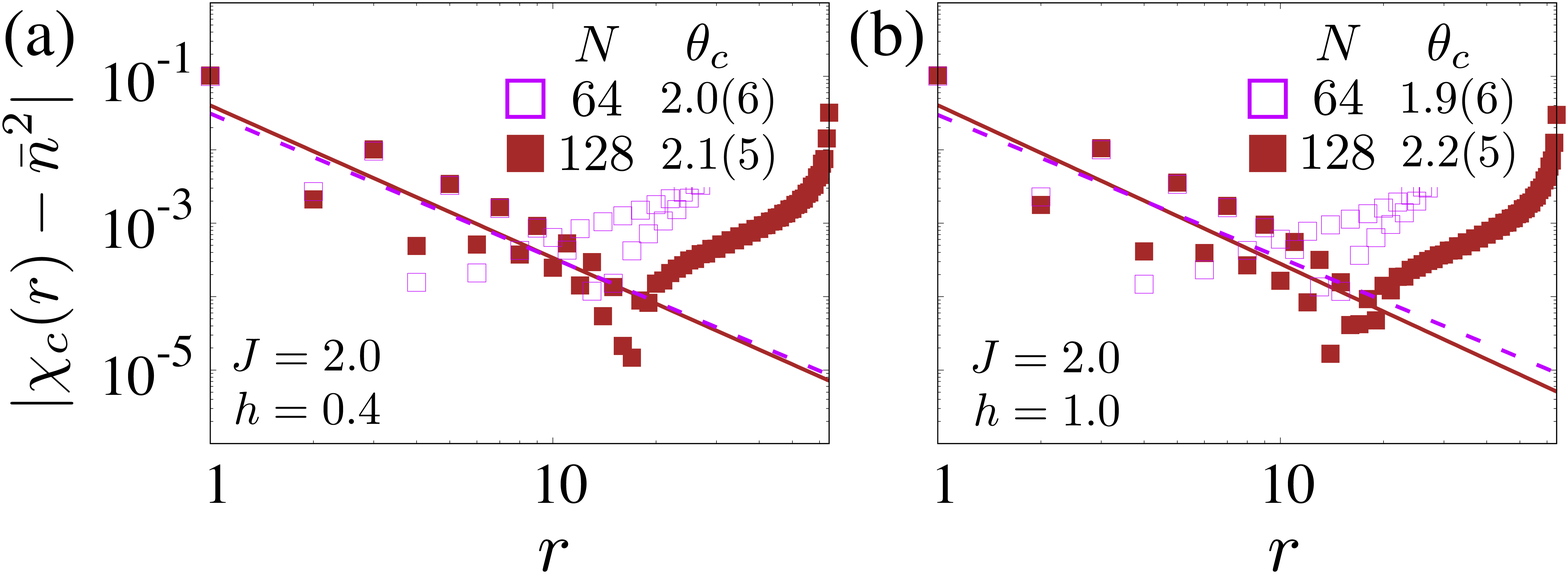}
		\caption{(Color online) 
		$r$ dependence of the charge correlation function $|\chi_{c}(r)-\bar{n}^{2}|$ for $N=64$ (open symbols), 128 (filled symbols), $n_{c}=0.5$ with the cutoff $m=200$ in the KP phase: (a) $(J, h)=(2.0, 0.4)$ and (b) $(J, h)=(2.0, 1.0)$.
		$\bar{n}^{2}$ is the average value of $\chi_{c}(r)$ from $r=20$ to $25$ for $N=64$ and $r=31$ to $40$ for $N=128$.
		The (dashed) lines represent the fit by $Ar^{-\theta_{c}}$ for $N=128\;(64)$ in the range $1 \leq r \leq 10$.
		}
		\label{J20h04_10_N0_CF_e64_32}
	\end{center}
\end{figure}

Figure \ref{J20h04_10_N0_CF_e64_32} shows the $r$ dependence of 
the charge correlation function 
$\chi_c(r)$, which is defined as 
\begin{align}
\chi_c(r):=\braket{\hat{n}_{r+N/2}\hat{n}_{N/2}}.
\end{align}
 It decays to a constant value $\bar{n}^2$ up to $r\sim 10$, and for larger $r$, the data are masked by the boundary effects. 
Apart from this boundary effect, it shows power-law behavior as $\chi_c(r)-\bar{n}^2\sim1/r^{\theta_c}$. Note that $\bar{n}^2$ is a fitting parameter here and $\bar{n}^2 \ne n_c^2$.
Thus, in the present numerical calculations, there is no
evidence \tred{of} the charge density wave (CDW) in the KP phase for $n_{c}=0.5.$

\begin{figure}[t]
	\begin{center}
		\includegraphics[scale=0.25]{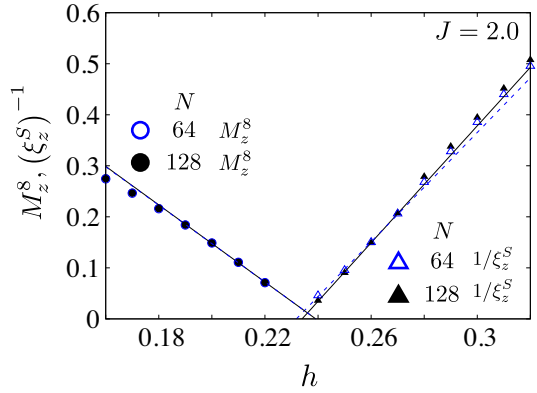}
		\caption{(Color online) 
		$h$ dependence of $M_z^{8}$ and $1/\xi_{z}^{S}$ for $J=2.0$ and $n_{c}=0.5$.
		Results are obtained from the linear extrapolation to the cutoff $1/m\rightarrow 0$ by using $m=100,\;150,\;200$, and $300$.
		The (dashed) lines represent the linear fits for $M_{z}^{8}$ and $1/\xi_{z}^{S}$ for $N=128\;(64)$.
		The fittings are carried out for \tred{$h=0.19\textendash0.22$} for $M_{z}^{8}$ and for \tred{$h=0.24\textendash0.27$} for $1/\xi_{z}^{S}$.
		$M_z\sim (h_{c}-h)^{1/8}$ and $\xi_z^S\sim (h-h_c)^{-1}$ suggest the \tred{2D} Ising universality class.
		}
		\label{critical_exponent}
	\end{center}
\end{figure}

Finally, we discuss \tred{the Ising-ordered states}, focusing on the universality class of 
the phase transition between the FM and \tred{KP} phases. 
The $h$ \tred{dependences} of the correlation length $\xi_z^S$ and the magnetization $M_z$ around
the phase boundary 
suggest that the transition belongs to the two-dimensional \tred{(2D)} Ising universality class.  
Figure \ref{critical_exponent} shows the $h$ dependence of $M_{z}$ and $\xi^S_z$ of $\chi^{S}_{z}(r)$ for $J = 2.0$. 
One can see that 
 \begin{equation}
M_{z}\simeq (h_c-h)^{\frac{1}{8}},\; \xi^S_z\simeq (h-h_c)^{-1},\; {\rm with}\; h_c\sim0.23\:{\rm -}\:0.24.
 \end{equation}
These exponents correspond to  $\beta = 1/8$ and $\nu = 1$ in the \tred{2D-Ising} class, 
which is consistent with the fact that the system possesses the Ising anisotropy.
The critical field \tred{$h_{c}\sim0.23$--$0.24$} is consistent with the
 metamagnetic field observed in the magnetization  in Fig. \ref{J20_mean_Sx1_Sx0_e64_FMTOKP}.

\begin{figure}[t]
	\begin{center}
		\includegraphics[scale=0.15]{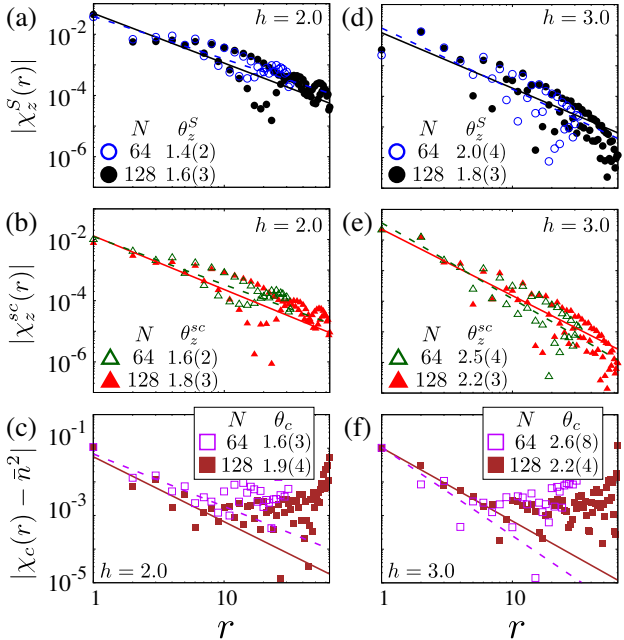}
		\caption{(Color online) 
		$r$ dependence of the spin-spin and \tred{charge} correlation functions for $N=64$ (open symbols), 128 (filled symbols), and $n_{c}=0.5$ with the cutoff $m=200$ in the TLL phase:
		(a) $|\chi_{z}^{S}(r)|$, (b) $|\chi_{z}^{sc}(r)|$, and (c) $|\chi_{c}(r)-\bar{n}^{2}|$ for $(J, h) = (2.0, 2.0)$, and
		(d) $|\chi_{z}^{S}(r)|$, (e) $|\chi_{z}^{sc}(r)|$, and (f) $|\chi_{c}(r)-\bar{n}^{2}|$ for $(J, h) = (2.0, 3.0)$.
		The (dashed) lines represent the fit by $Ar^{-\theta}$ for $N=128\;(64)$.
		The range of the fitting for $|\chi_{c}(r)-\bar{n}^{2}|$ is the same as that in Fig. \ref{J20h04_10_N0_CF_e64_32}.
		}
		\label{J20h30_Sz1_CF_Sz0_CF}
	\end{center}
\end{figure}

Let us close the discussion about correlation functions by examining them in the TLL phase. 
Figure \ref{J20h30_Sz1_CF_Sz0_CF} shows the $r$ dependence of $\chi^{S}_z(r)$, $\chi^{sc}_z(r)$, and $\chi_{c}(r)$ for $(J, h)=(2.0, 2.0)$ and $(2.0, 3.0)$.
The results indicate the power-law decay $\sim 1/r^{\theta}$, 
reflecting \tred{the fact} that the system is critical, 
although the numerical accuracy is \tred{insufficient} for the precise determination of the exponent $\theta$. 
For this, we need more elaborate calculations, \tred{which will be tackled as a future problem}.

\subsection{Friedel \tred{oscillations}}

\tred{Here, we} analyze the spatial dependence of the expectation values: $M_{j}^{x}:=\braket{\hat{S}^x_j}$ and $m_{j}^{x}:=\braket{\hat{s}^x_j}$.
Their spatial dependence \tred{is affected by the presence of the edges. $M_{j}^{x}$ and $m_{j}^{x}$ show characteristic oscillations and decay\cite{ODKL_Friedel}}, 
which reflect the ground state of the system.
This is known as Friedel oscillations\cite{FO_1}.

\begin{figure}[t]
	\begin{center}
		\includegraphics[scale=0.28]{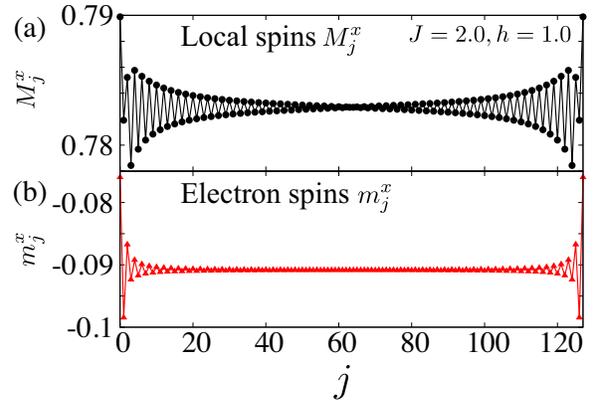}
		\caption{(Color online) 
Spatial dependence of the magnetizations (a) $M_{j}^{x}$ and (b) $m_{j}^{x}$ for $N=128$, $J=2.0$, $h=1.0$, and $n_{c}=0.5$ with the cutoff $m=200$ in the KP phase.
The oscillations decay exponentially to $j=64$ from both \tred{the} left and right edges. The correlation lengths $\xi_x^S$ and $\xi_x^c$ are strongly \tred{size-dependent}. See the main text.
}
		\label{J20h10_Sx1_Sx0_e64}
	\end{center}
\end{figure}

Figures \ref{J20h10_Sx1_Sx0_e64}(a) and \ref{J20h10_Sx1_Sx0_e64}(b) show the spatial dependence of $M_{j}^{x}$ and $m_{j}^{x}$, respectively, for $J = 2.0$ and $h = 1.0$ in the KP phase.
The oscillations of these local magnetizations seem to decay exponentially toward the middle of the chain.
By fitting the data \tred{to the} form $c_1+c_2 \exp(-j/\xi_x)$ for $20<j<40$, 
we can estimate the correlation length $\xi_x$.
We find \tred{that} $\xi^S_x=10.20$ for $M_{j}^{x}$ and $\xi^{sc}_x=8.09$ for $m_{j}^{x}$ for $N=64$ and $\xi^S_x=19.46$ and $\xi^{sc}_x=16.95$ for $N=128$ \tred{about} twice the length of that for $N=64$.
This suggests that the system size is much smaller than true $\xi_{x}^{S,sc}$. \tred{Because} the system is metallic, it is natural to expect that the longitudinal fluctuation is not gapped, i.e., $\xi^{S,sc}_{x}\rightarrow\infty$ in the thermodynamic limit.

\begin{figure}[t]
	\begin{center}
		\includegraphics[scale=0.28]{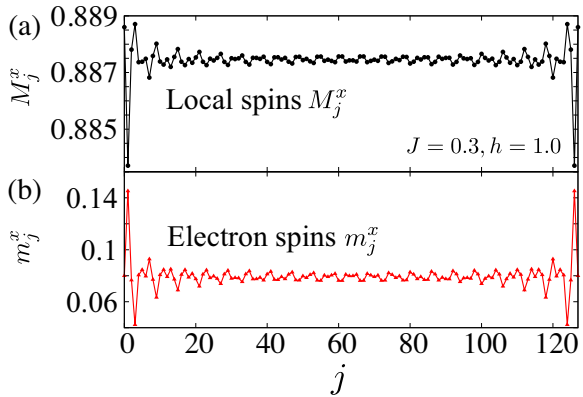}
		\caption{(Color online) 
		Spatial dependence of the magnetizations (a) $M_{j}^{x}$ and (b) $m_{j}^{x}$ for $N=128$, $J=0.3$, $h=1.0$, and $n_{c}=0.5$ with the cutoff $m=200$ in the TLL phase.
		The amplitude of the oscillations is so small that it is impossible to judge whether they show exponential or \tred{power-law} decay.
		}
		\label{J20h30_Sx1_Sx0_e64}
	\end{center}
\end{figure}

\begin{figure}[t]
	\begin{center}
		\includegraphics[scale=0.25]{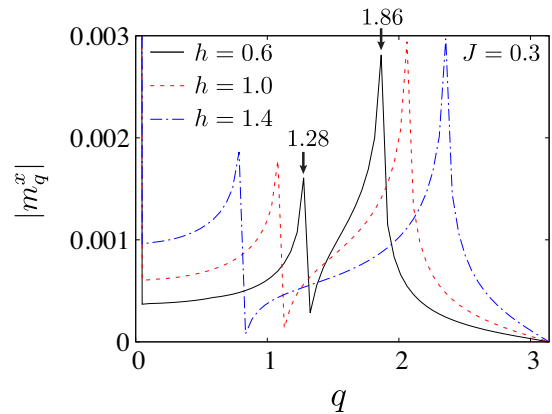}
		\caption{(Color online) 
		Wavenumber $q$ dependence of $|m^{x}_{q}|$ for $N=128$, $J=0.3$, and $n_{c}=0.5$ with the cutoff $m=200$. 
		Data for three sets of the magnetic fields ($h=0.6$, $1.0$, and $1.4$) are shown for clarity and the two peaks in each $h$ correspond to $2k_{F\pm}$.
		Note that there is a large $q=0$ component since $m_x$ is induced by $h$.
		}
		\label{fourier_components}
	\end{center}
\end{figure}

When the system is critical, the profile of the oscillation \tred{markedly} changes.
Figures \ref{J20h30_Sx1_Sx0_e64}(a) and \ref{J20h30_Sx1_Sx0_e64}(b) show $M_{j}^{x}$ and $m_{j}^{x}$, respectively, for $J = 2.0$ and $h = 3.0$, 
where the system is in the TLL phase. 
The long-range oscillation, which is one of the characteristics of \tred{a TLL}, is 
induced by the presence of the edges. The dominant components of this oscillation correspond to two kinds of Fermi wavenumbers $k_{{\rm F}\pm}$, 
where + $(-)$ indicates the upper (lower) conduction electron band.
For $J = 0$, they are given as
\begin{align}
k_{\rm F\pm}^{0}(h) = \frac{\pi n_{c}}{2}\mp\arcsin\frac{h}{4\sin\frac{\pi n_{c}}{2}}.
\end{align}
Suppose the Fermi surface (point) is ``small'' as 
expected  for \tred{the} $S=1$ Kondo lattice model\cite{Yamanaka,Oshikawa},
$k_{{\rm F}\pm}$ is identical to $k_{\rm F\pm}^{0}$.
Absolute values of the Fourier transform $m_{q}^{x}$ of $m_{j}^{x}$ for $h=0.6$, $1.0$, and $1.4$ with fixed $J=0.3$ are shown in Fig. \ref{fourier_components}.
There are two peaks for all the three \tred{$h$}.
For example, for $h=0.6$, the two peaks \tred{are located} at $q_1 = 1.28$ and $q_2 = 1.86$,
which are very close to $2k_{\rm F+}^{0}(h_{\rm{eff}})=1.31$ and $2k_{\rm F-}^{0}(h_{\rm{eff}})=1.83$. 
Here, $h_{\rm{eff}}=h-JM_{x}(h=0.6)=0.37$ is the effective field for the conduction electrons.
Similarly, for $h=1.0$ and $1.4$, there are two peaks corresponding to $2k_{\rm F\pm}^{0}(h_{\rm{eff}})$.
This is consistent with the ``small Fermi surface''. 
For larger $J$, the analysis based on $h_{\rm{eff}}$ does not work and the spin correlations are not governed by the Fermi points.

\subsection{Filling dependence}\label{sec:3.5}
So far, we have discussed the case for $n_c=0.5$. 
In this subsection, we will briefly show the results for other fillings $n_c=0.25,0.75$, and $1.0$,
since most of the phases appearing \tred{for} $n_c\ne 0.5$ are similar to those appearing for $n_c=0.5$.
One qualitatively different point from \tred{the case of} $n_c=0.5$ is \tred{that a CDW state appears in the KP and AFM phases for $n_c=0.75$.}

\begin{figure}[t]
	\begin{center}
	 \includegraphics[scale=0.25]{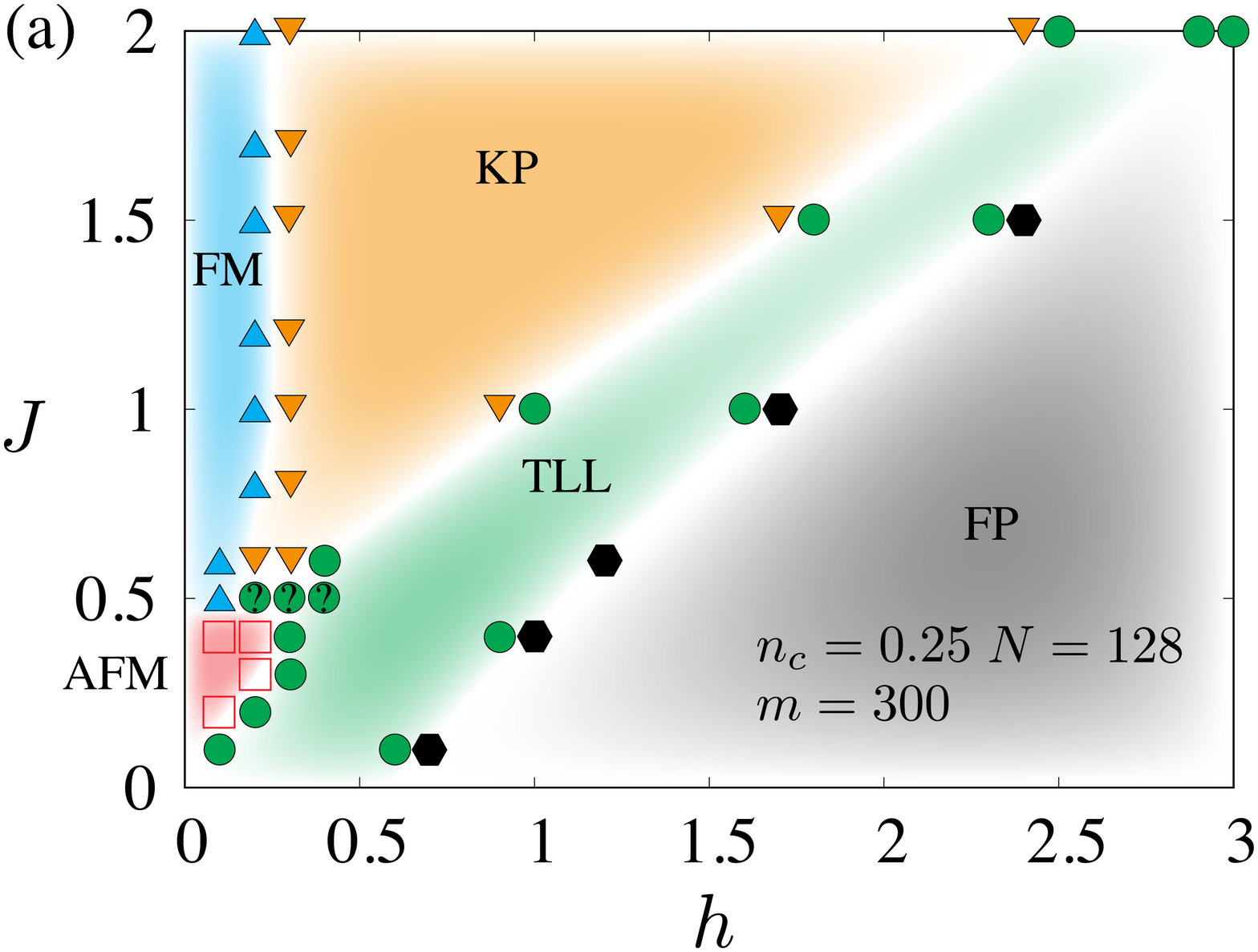}
	\end{center}
	
	\begin{center}
	 \includegraphics[scale=0.25]{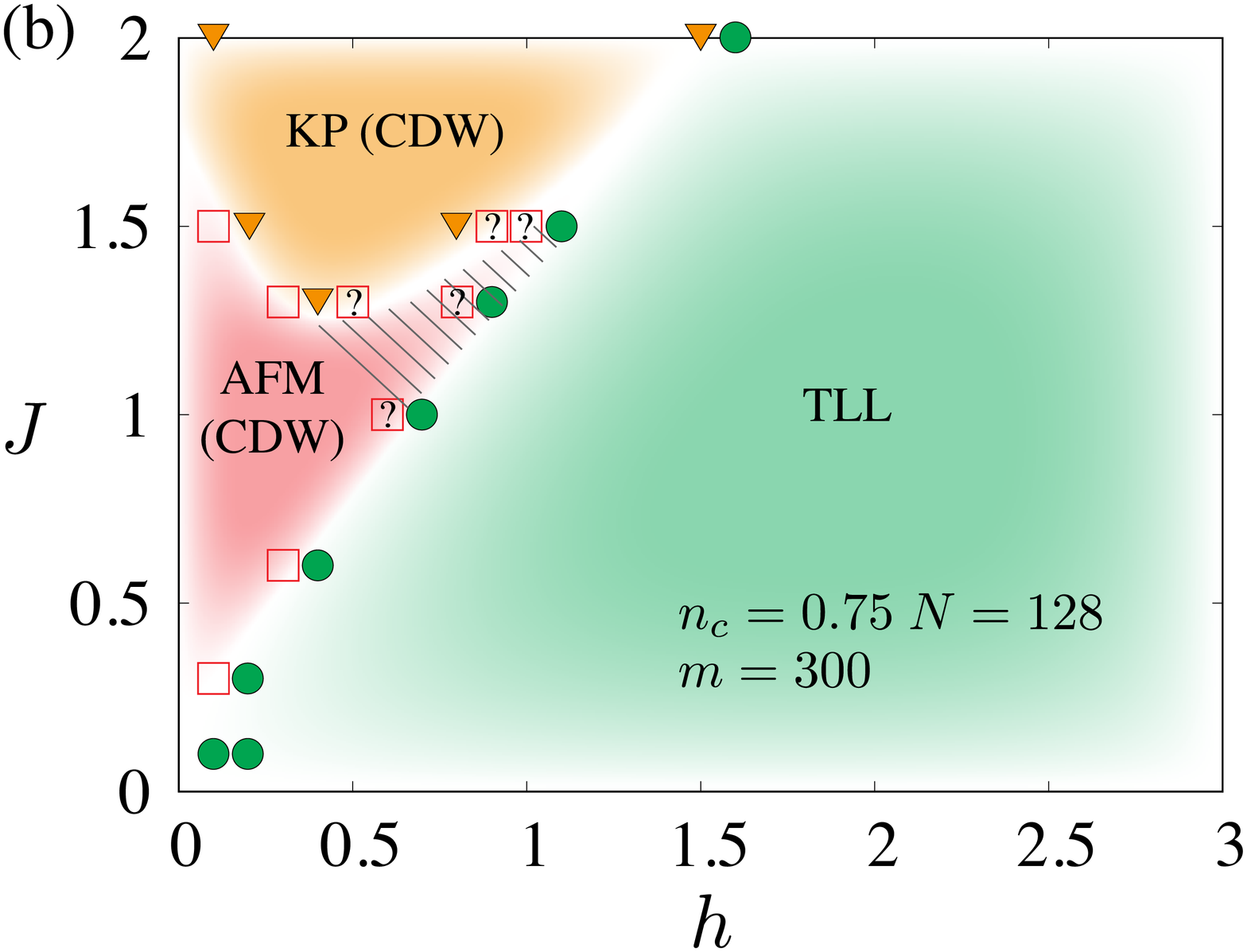}
	\end{center}
	
	\begin{center}
	 \includegraphics[scale=0.25]{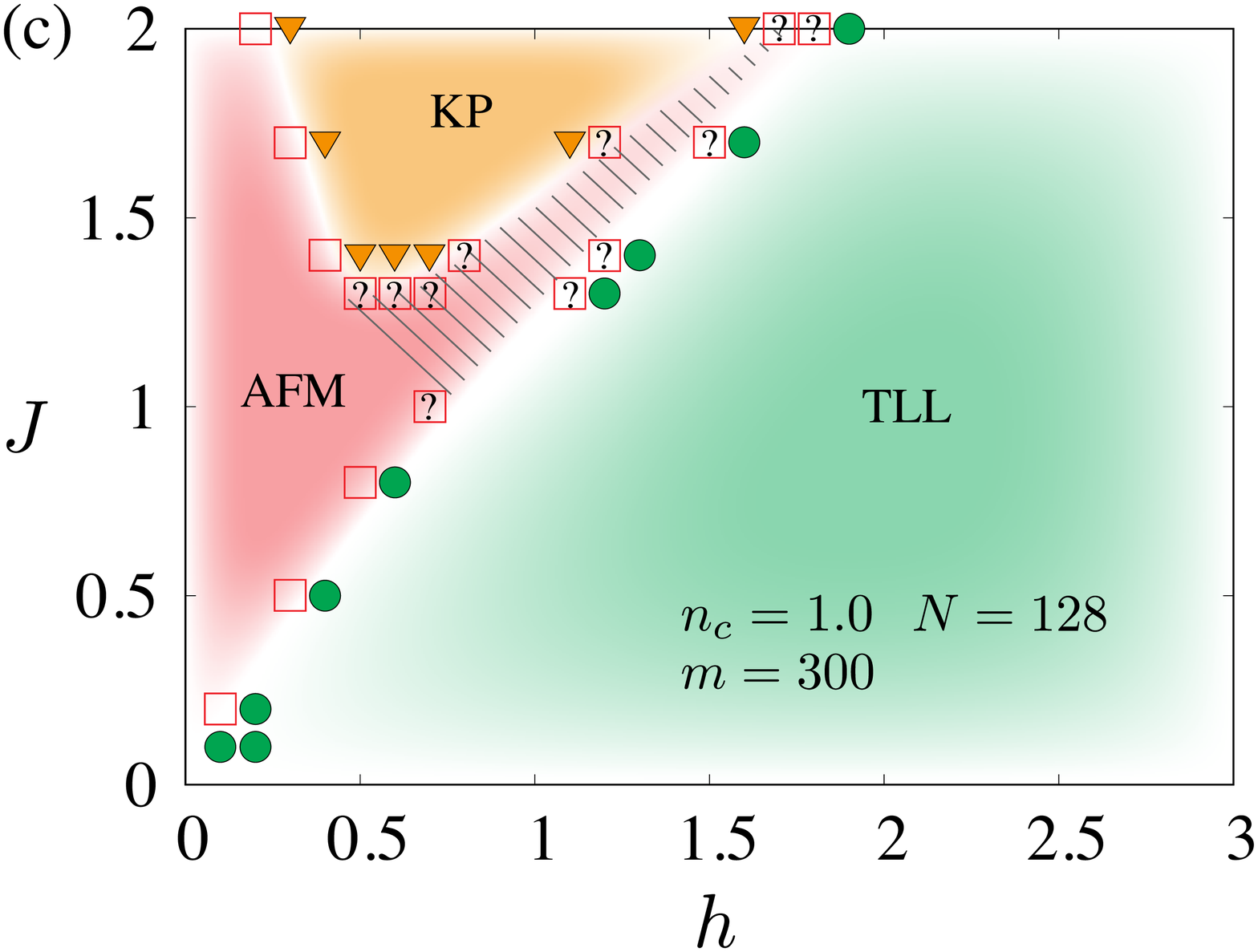}
	\caption{(Color online) Ground-state $h$-$J$ phase diagram for (a) $n_{c} = 0.25$, (b) $n_{c} = 0.75$, and (c) $n_{c} = 1.0$. 
	The phases are determined by analyzing the behavior of the correlation functions for $N = 64$ and $128$ with the cutoff $m = 300$. 
	\tred{The} symbols represent points where numerical calculations have been \tred{carried out}, and their type and color distinguish the ground state. 
	For clarity, only \tred{the} symbols near phase boundaries are indicated. 
	Symbols with ``?'' near phase boundaries and the shaded area indicate that the ground state is not identified \tred{owing} to the finite size effects and/or insufficient cutoff $m$. 
	See the explanation in the main text. 
	For (b) and (c), the FP phase \tred{is located at} $h > 3$.
	}
	\label{other_phase}
	\end{center}
\end{figure}

Figures \ref{other_phase}(a)--\ref{other_phase}(c) 
show the $h$-$J$ phase diagrams for $n_c = 0.25$, $0.75$, and $1.0$, respectively. 
For the smallest filling $n_{c}=0.25$, the FM phase appears for \tred{most} of the Ising-ordered phases and the KP and \tred{FP} phases occupy \tred{a} wide range of the phase diagram as shown in Fig. \ref{other_phase}(a).
For the larger fillings, the FM phase disappears and the AFM phase extends \tred{to} larger $h$ compared with the smaller fillings.
For $n_c=0.75$ and $1.0$ [Figs. \ref{other_phase}(b) and \ref{other_phase}(c)], however, there is ambiguity in the long-distance behavior of $\chi^{S,sc}_z(r)$ for the shaded areas; 
$\chi^{S,sc}_z(r)$ for large $r$ seems to still decay slightly, 
while the oscillation is similar to that in the AFM phase. 
We have examined \tred{$\chi^{S,sc}_z(r)$} for the cutoff $m$ up to $500$, but this behavior persists. 
Thus, the presence of the AFM phase in the shaded areas \tred{is considered to be an} artifact due to the finite size effects and/or insufficient cutoff number near the phase boundary, 
although it \tred{would be} interesting if the reentrant phase diagram as a function of $h$ \tred{was} realized.
When $J=0$ and $n_{c}=0.75$, the FP phase appears for $h\gtrsim3.62$ and for $n_{c}=1.0$, $h>4$.

\begin{figure}[t]
	\begin{center}
		\includegraphics[scale=0.25]{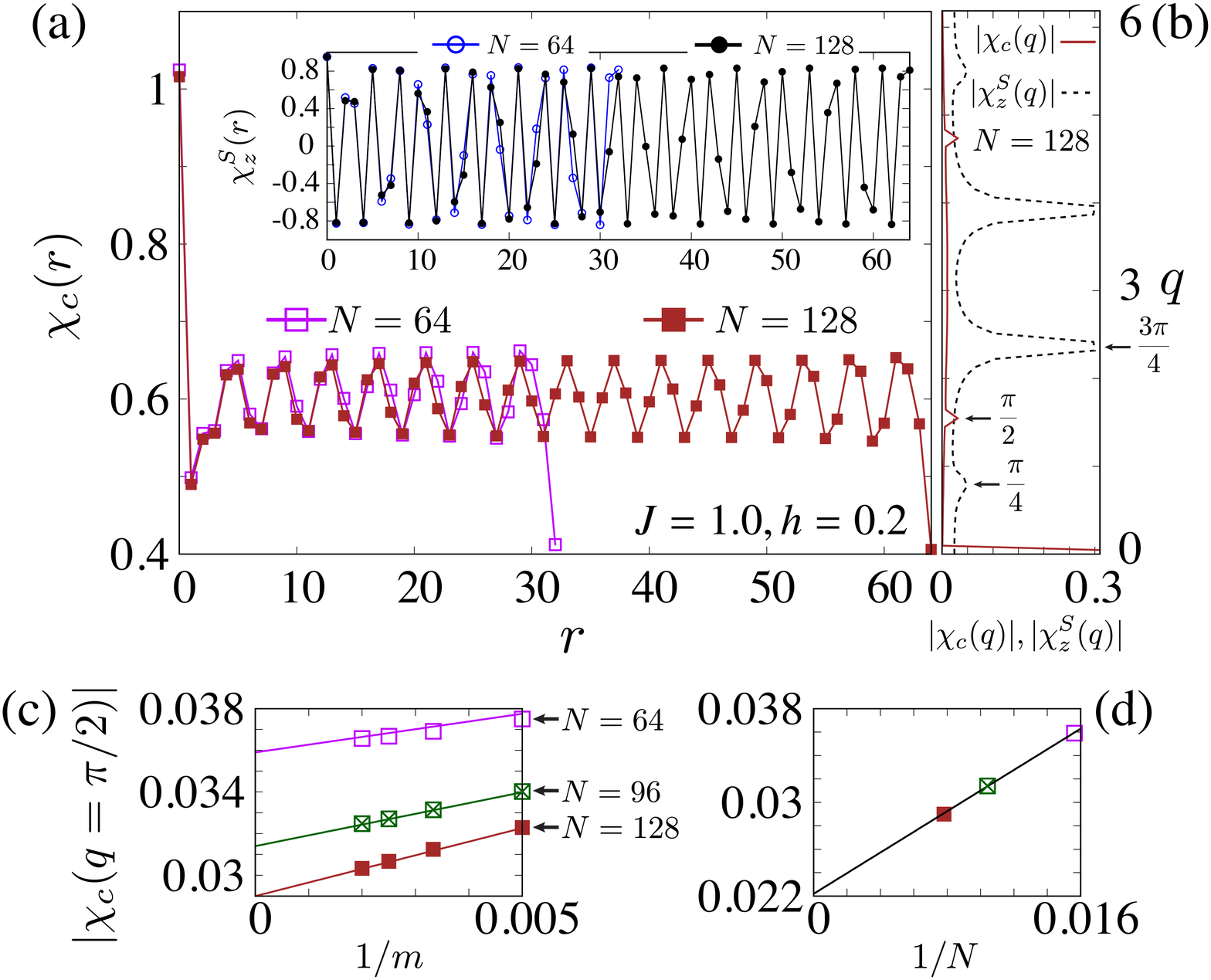}
		\caption{(Color online) 
		Comparison between the charge and spin correlations in the AFM phase for $n_{c}=0.75$, $J=1.0$, and $h=0.2$ with the cutoff $m=400$. 
		(a) Spatial dependence of $\chi_{c}(r)$. Inset: $\chi_{z}^{S}(r)$ as a function of $r$. 
		(b) Fourier \tred{transforms} $\chi_{c}(q)$ and $\chi_{z}^{S}(q)$ for $\chi_{c}(r)$ and $\chi_{z}^{S}(r)$, respectively, for $N=128$.
		(c) \tred{Cutoff} $m$ dependence of $\chi_{c}(q=\pi/2)$. The lines represent the linear fits by using the \tred{data} for $m=200, 300, 400$, and $500$.
		(d) $N$ dependence of $\chi_{c}(q=\pi/2)$. 
		\tred{The} symbols correspond to $1/m\rightarrow 0$ values for each $N$ in (c). 
		The line represents the linear fit by using the \tred{data} for $N=64, 96$, and $128$.}
		\label{J10h02N0_Sz1_CF}
	\end{center}
\end{figure}

As mentioned above, the CDW coexists with the AFM and \tred{KP} states for $n_c=0.75$ in Fig. \ref{other_phase}(b).
Figure \ref{J10h02N0_Sz1_CF}(a) shows the $r$ dependence of the charge correlation function $\chi_{c}(r)$ for $J=1.0$ and $h=0.2$ in the AFM phase.
One can see the oscillation with the wavenumber $q=\pi/2$, 
which corresponds to $4k_{\rm F}^{0}=\pi/2$ mod $2\pi$ for $h=0$.
This is verified by the direct Fourier transform as shown in Fig. \ref{J10h02N0_Sz1_CF}(b).
A similar CDW with $q=4k_{\rm F}^{0}$ appears also for $n_c=0.8$ and 0.85 (not shown).
Figures \ref{J10h02N0_Sz1_CF}(c) and \ref{J10h02N0_Sz1_CF}(d) show the cutoff $m$ and the system size $N$ dependence of the dominant Fourier component for $q=\pi/2$.
Extrapolating them to $1/m\rightarrow 0$ for each $N$ and then to $1/N\rightarrow 0$, 
we obtain a finite value $\chi_c(q=\pi/2)\sim 0.022$.
In contrast to $\chi_c$, the spin correlations $\chi^{S}_{z}$ and $\chi^{sc}_{z}$ are governed by the $2k_{\rm F}$ fluctuations as discussed before.
The inset of Fig. \ref{J10h02N0_Sz1_CF}(a) shows  $\chi^{S}_{z}(r)$ and there are complex modulations. 
The main Fourier mode is that for $q=3\pi/4$ as shown in Fig. \ref{J10h02N0_Sz1_CF}(b), which is \tred{the} $2k_{\rm F}^{0}$ mode.
There is another peak for $q=\pi/4$, which corresponds to $6k_{\rm F}^{0}=\pi/4$ mod $2\pi$.
This is considered to be the coupling mode of the CDW with $q=4k_{\rm F}^{0}$ and the AFM state with $q=2k_{\rm F}^{0}$.
In the one-dimensional spin-1/2 Kondo lattice model, the spin density oscillations induced by the presence of \tred{a} boundary have \tred{a} two-peak structure corresponding to \tred{the} $2k_{\rm F}^{0}$ mode 
and the coupling mode between the spin- and \tred{charge-density} oscillations\cite{ODKL_Friedel}. 
The amplitude of the latter decreases with increasing \tred{system} size. 
Thus, also in our calculations, there is a possibility that the structure corresponding to $6k_{\rm F}^{0}$ disappears in the thermodynamic limit.
In two- and infinite-dimensional spin-1/2 Kondo lattice models, 
CDWs are reported at finite temperatures for $n_{c}\simeq0.5$.\cite{2d_KL_CDW,KL_CDW}
In our DMRG study, however, there is no evidence for the CDW for $n_{c}=0.5$. 
More detailed and sophisticated investigations\cite{Shibata_KLBoundary,Katsura} are needed 
to gain a deep understanding about \tred{the} emergence of such CDW states, 
since the amplitudes for the CDW obtained are relatively small.

\section{\tred{Discussion} and Summary}
\subsection{Comparison with the experimental results \tred{for} URhGe}
In this section, we compare our results with the experimental results \tred{for} URhGe.
We have studied the simplified one-dimensional model \tred{[Eq. (\ref{eq_ham})]}, keeping the \tred{FM} superconductor URhGe in mind.
In URhGe, as the external magnetic field $H$ is applied to the hard axis (b-axis), the SC disappears at \tred{$H\sim2\;\rm{T}$},
but it reappears for $H=9\sim13\;{\rm T}$.
In particular, the transition temperature $T_{\rm{sc}}$ is highly enhanced and has a maximum value \tred{of} 0.42 K  for $H=H_{R}\sim12\;\rm{T}$.
This enhancement in $T_{\rm sc}$ is closely related to the spin reorientation at $H_{R}$ and also to \tred{the} metamagnetism \cite{URhGe_metamag} \tred{originating} from
the tricriticality\cite{URhGe_TCP_TEP,URhGe_TCP_NMR}. 

In our model, \tred{metamagnetic enhancement of $M_{x}$ appears} around $h_{c}$
as shown in Fig. \ref{J20_mean_Sx1_Sx0_e64_FMTOKP}. This is in sharp contrast to the spin-wave analysis in Ref. \citen{URhGe_theor_XXZ}, where the longitudinal magnetization is essentially given by 
that for the mean-field approximation. In \tred{the DMRG analysis}, quantum fluctuations are fully taken into account, \tred{which} is necessary for the metamagnetic behavior \tred{shown} in Fig. \ref{J20_mean_Sx1_Sx0_e64_FMTOKP}.
In the experimental situation, the tricritical point \tred{is located at} \tred{$T\sim 2$--$4\;{\rm K}$}\cite{URhGe_TCP_TEP,URhGe_TCP_NMR} $> T_{\rm sc}$. 
Thus, the SC dome should be across the first-order line. One might wonder why 
the \tred{SC} is induced by the first-order transition since the fluctuations
are usually small. However, the NMR experiment\cite{URhGe_fluc1,URhGe_fluc2} has revealed that there is 
strong fluctuation even in such a situation \tred{owing} to the proximity to the tricritical point.
As analyzed in \tred{detail} in Sect. 3.3, the phase transition between the Ising FM and \tred{KP} phases is \tred{second-order} with the \tred{2D} Ising class.
Thus, there is no tricritical point in our model. Nevertheless, it \tred{will be} interesting to examine whether SC fluctuations 
are \tred{enhanced and} also which channels of SC are favored near the phase boundary in future studies.
\tred{Note that the TLL phase is always} next to the FP phase \tred{on} the low-field side, and also the high-field 
boundary of the FM phase faces the KP phase, which represents strong Kondo correlations, 
for \tred{the} parameter sets we have examined. 
Thus, various phases compete near the FM phase under transverse 
fields in the present Kondo lattice model. Generally, in such a situation, there are various fluctuations and it is 
interesting to examine the possibility of enhanced superconducting fluctuations. We expect \tred{that} such \tred{a} situation can capture the physics of URhGe under \tred{transverse} fields.

\subsection{Summary}
We have analyzed the $S = 1$ one-dimensional Kondo lattice model with a uniaxial anisotropy under a transverse field $h$ by using \tred{the} DMRG, which is a simplified model for \tred{the} U-based \tred{FM} \tred{superconductor} URhGe and related compounds. 
We have constructed the ground-state phase diagram as functions of the magnetic field $h$ and the Kondo exchange coupling $J$ for several conduction electron fillings. The phase diagram includes various phases such as 
the Ising-ordered FM and AFM phases, \tred{the} TLL phase, and the fully polarized (FP) phase. In \tred{addition}, a (spin) gapped phase \tred{appears} in moderate $h$ between the FM and \tred{TLL} states, where the local and conduction electron spins are tightly 
bound antiferromagnetically and the magnetization there shows a plateau-like $h$ dependence. We have dubbed this \tred{the} Kondo plateau (KP) phase. We have also discussed the metamagnetic behavior when the FM state is destabilized by applying the transverse field $h$ and found a steep increase in the magnetization near the critical field for the FM phase. This behavior is consistent with the magnetization data for URhGe. Examining the superconducting fluctuation near the critical field is an interesting future direction and is now in progress.

\section*{Acknowledgment}
This work was supported by a Grant-in-Aid for Scientific Research (Grant Nos. 16H01079 \tred{and} 18K03522) from the Japan Society for the Promotion of Science.


\begin{thebibliography}{99} 
\bibitem{UGe2_FMS} S. S. Saxena, P. Agarwal, K. Ahilan, F. M. Grosche, R. K. W. Haselwimmer, M. J. Steiner, E. Pugh, I. R. Walker, S. R. Julian, P. Monthoux, G. G. Lonzarich, A. Huxley, I. Sheikin, D. Braithwaite, and J. Flouquet, Nature {\bf406}, 587 (2000).
\bibitem{URhGe_FMS} D. Aoki, A. Huxley, E. Ressouche, D. Braithwaite, J. Flouquet, J. P. Brison, E. Lhotel, and C. Paulsen,  Nature {\bf413}, 613 (2001).
\bibitem{UIr_FMS} T. Akazawa, H. Hidaka, T. Fujiwara, T. C. Kobayashi, E. Yamamoto, Y. Haga, R. Settai, and Y. \={O}nuki, J. Phys.: Condens. Matter {\bf16}, L29 (2004).
\bibitem{UCoGe_FMS} N. T. Huy, A. Gasparini, D. E. de Nijs, Y. Huang, J. C. P. Klaasse, T. Gortenmulder, A. de Visser, A. Hamann, T. G\"{o}rlach, and H. v. L\"{o}hneysen, Phys. Rev. Lett. {\bf99}, 067006 (2007).
\bibitem{URhGe_RSC} F. L\'{e}vy, I. Sheikin, B. Grenier, and A. D. Huxley, Science {\bf309}, 1343 (2005).
\bibitem{URhGe_HTphadia} A. Miyake, D. Aoki, and J. Flouquet, J. Phys. Soc. Jpn. {\bf77}, 094709 (2008).
\bibitem{UCoGe_PTphadia} E. Hassinger, D. Aoki, G. Knebel, and J. Flouquet, J. Phys. Soc. Jpn. {\bf77}, 073703 (2008).
\bibitem{UCoGe_Hc2} D. Aoki, T. D. Matsuda, V. Taufour, E. Hassinger, G. Knebel, and J. Flouquet, J. Phys. Soc. Jpn. {\bf78}, 113709 (2009).
\bibitem{URhGe_fluc1} Y. Tokunaga, D. Aoki, H. Mayaffre, S. Kr\"{a}mer, M.-H. Julien, C. Berthier, M. Horvati\'c, H. Sakai, S. Kambe, and S. Araki, Phys. Rev. Lett. {\bf114}, 216401 (2015).
\bibitem{URhGe_fluc2} Y. Tokunaga, D. Aoki, H. Mayaffre, S. Kr\"{a}mer, M.-H. Julien, C. Berthier, M. Horvati\'{c}, H. Sakai, T. Hattori, S. Kambe, and S. Araki, Phys. Rev. B {\bf93}, 201112 (2016).
\bibitem{UCoGe_fluc1} T. Hattori, Y. Ihara, Y. Nakai, K. Ishida, Y. Tada, S. Fujimoto, N. Kawakami, E. Osaki, K. Deguchi, N. K. Sato, and I. Satoh, Phys. Rev. Lett. {\bf108}, 066403 (2012).
\bibitem{UCocGe_fluc2} T. Hattori, K. Karube, K. Ishida, K. Deguchi, N. K. Sato, and T. Yamamura, J. Phys. Soc. Jpn. {\bf83}, 073708 (2014).
\bibitem{QPT_Book} S. Sachdev, {\it Quantum Phase Transitions} (Cambridge University Press, Cambridge, MA, 1999) p. 39.
\bibitem{URhGe_TCP_TEP} A. Gourgout, A. Pourret, G. Knebel, D. Aoki, G. Seyfarth, and J. Flouquet, Phys. Rev. Lett. {\bf117}, 046401 (2016).
\bibitem{URhGe_TCP_NMR} H. Kotegawa, K. Fukumoto, T. Toyama, H. Tou, H. Harima, A. Harada, Y. Kitaoka, Y. Haga, E. Yamamoto, Y.  \={O}nuki, K. M. Itoh, and E. E. Haller, J. Phys. Soc. Jpn. {\bf84}, 054710 (2015).
\bibitem{URhGe_metamag} F. Hardy, D. Aoki, C. Meingast, P. Schweiss, P. Burger, H. v. L\"{o}hneysen, and J. Flouquet, Phys. Rev. {\bf83}, 195107 (2011).
\bibitem{URhGe_Mineev1} V. P. Mineev, Phys. Rev. B {\bf83}, 064515 (2011).
\bibitem{URhGe_theor_XXZ} K. Hattori and H. Tsunetsugu, Phys. Rev. B {\bf87}, 064501 (2013).
\bibitem{Christopher} C. L\"{o}rscher, J. Zhang, Q. Gu, and R. A. Klemm, Phys. Rev. B {\bf88}, 024504 (2013).
\bibitem{URhGe_Mineev2} V. P. Mineev, Phys. Rev. B {\bf90}, 064506 (2014).
\bibitem{URhGe_Mineev3} V. P. Mineev, Phys. Rev. B {\bf91}, 014506 (2015).
\bibitem{URhGe_Mineev4} V. P. Mineev, Phys. Rev. B {\bf96}, 104501 (2017).
\bibitem{UCoGe_theor} Y. Tada, S. Takayoshi, and S. Fujimoto, Phys. Rev. B {\bf93}, 174512 (2016).
\bibitem{Lifshitz_trransition} Y. Sherkunov, A. V. Chubukov, and J. J. Betouras, Phys. Rev. Lett. {\bf121}, 097001 (2018).
\bibitem{LineNode_UCoGe} T. Nomoto and H. Ikeda, J. Phys. Soc. Jpn. {\bf86}, 023703 (2017).
\bibitem{Topological_SC_UCoGe} A. Daido, T. Yoshida, and Y. Yanase, arXiv:1803.07786.
\bibitem{ODKL_GS} H. Tsunetsugu, M. Sigrist, and K. Ueda, Rev. Mod. Phys. {\bf69}, 809 (1997).
\bibitem{Ueda_KLFerro} K. Ueda, T. Nishino, and H. Tsunetsugu, Phys. Rev. B {\bf50}, 612 (1994).
\bibitem{ODKL_Review} N. Shibata and K Ueda, J. Phys.: Condens. Matter {\bf11}, 4289 (1999).
\bibitem{Assad} F. F. Assaad, Phys. Rev. Lett. {\bf83}, 796 (1999).
\bibitem{Otsuki} J. Otsuki, H. Kusunose, and Y. Kuramoto, Phys. Rev. Lett. {\bf102}, 017202 (2009).
\bibitem{Smerat_KLQuasi} S. Smerat, U. Schollw\"{o}ck, I. P. McCulloch, and H. Schoeller, Phys. Rev. B {\bf79}, 235107 (2009).
\bibitem{Troyer_KLFerro} M. Troyer and D. W\"{u}rtz, Phys. Rev. B {\bf47}, 2886 (1993).
\bibitem{Tsunetsugu_KLFerro} H. Tsunetsugu, M. Sigrist, and K. Ueda, Phys. Rev. B {\bf47}, 8345 (1993).
\bibitem{Honner_Phase} G. Honner and M. Gul\'{a}csi, Phys. Rev. Lett. {\bf78}, 2180 (1997).
\bibitem{Honner_Phase2} G. Honner and M. Gul\'{a}csi, Phys. Rev. B {\bf58}, 2662 (1998).
\bibitem{McCulloch_KLFerro} I. P. Mcculloch, A. Juozapavicius, A. Rosengren, and M. Gul\'{a}csi, Philos. Mag. Lett. {\bf81}, 869 (2001).
\bibitem{McCulloch_KLFerro2} I. P. McCulloch, A. Juozapavicius, A. Rosengren, and M. Gulacsi, Phys. Rev. B {\bf65}, 052410 (2002).
\bibitem{Juozapavicius_KLFerro} A. Juozapavicius, I. P. McCulloch, M. Gulacsi, and A. Rosengren, Philos. Mag. B {\bf82}, 1211 (2002).
\bibitem{Garcia_KL_HL} D. J. Garcia, K. Hallberg, B. Alascio, and M. Avignon, Phys. Rev. Lett. {\bf93}, 177204 (2004).
\bibitem{Basylko_KLED} S. A. Basylko, P. H. Lundow, and A. Rosengren, Phys. Rev. B {\bf77}, 073103 (2008).
\bibitem{Peters_KLFerro} R. Peters and N. Kawakami, Phys. Rev. B {\bf86}, 165107 (2012).
\bibitem{ODKL_S1_1} C. Thomas, A. S. da Rosa Sim\~{o}es, J. R. Iglesias, C. Lacroix, N. B. Perkins, and B. Coqblin, Phys. Rev. B {\bf83}, 014415 (2011).
\bibitem{ODKL_S1_2} C. Thomas, A. S. da Rosa Sim\~{o}es, C. Lacroix, J. R. Iglesias, and B. Coqblin, J. Magn. Magn. Mater. {\bf372}, 247 (2014).
\bibitem{koga} M. Koga, G. Zar\'{a}nd, and D. L. Cox, Phys. Rev. Lett. {\bf83}, 2421 (1999).
\bibitem{DMRG} S. R. White, Phys. Rev. Lett. {\bf69}, 2863 (1992); Phys. Rev. B {\bf48}, 10345 (1993).
\bibitem{double-exchange} E. Dagotto, T. Hotta, and A. Moreo, Phys. Rep. {\bf344}, 1 (2001).
\bibitem{RKKY1} M. A. Ruderman and C. Kittel, Phys. Rev. {\bf96}, 99 (1954).
\bibitem{RKKY2} T. Kasuya, Prog. Theor. Phys. {\bf16}, 45 (1956).
\bibitem{RKKY3} K. Yosida, Phys. Rev. {\bf106}, 893 (1957).
\bibitem{ODKL_Friedel} N. Shibata, K. Ueda, T. Nishino, and C. Ishii, Phys. Rev. B {\bf54}, 13495 (1996).
\bibitem{FO_1} J. Friedel, Philos. Mag. {\bf43}, 153 (1952).
\bibitem{Yamanaka} M. Yamanaka, M. Oshikawa, and I. Affleck, Phys. Rev. Lett. {\bf79}, 1110 (1997).
\bibitem{Oshikawa} M. Oshikawa, Phys. Rev. Lett. {\bf84}, 3370 (2000).
\bibitem{2d_KL_CDW} T. Misawa, J. Yoshitake, and Y. Motome, Phys. Rev. Lett. {\bf110}, 246401 (2013).
\bibitem{KL_CDW} J. Otsuki, H. Kusunose, and Y. Kuramoto, J. Phys. Soc. Jpn. {\bf78}, 014702 (2009); J. Phys. Soc. Jpn. {\bf78}, 034719 (2009).
\bibitem{Shibata_KLBoundary} N. Shibata and C. Hotta, Phys. Rev. B {\bf84}, 115116 (2011).
\bibitem{Katsura} H. Katsura, J. Phys. A {\bf45}, 115003 (2012).
\end{thebibliography}
\end{document}